\documentstyle[12pt,aasms4]{article}

\begin{document}
\title{ Chemical Compositions of Four Metal-poor Giants  }
\author{
Sunetra Giridhar}
\affil{Indian Institute of Astrophysics;
Bangalore,  560034 India\\
giridhar@iiap.ernet.in}

\author{David L.\ Lambert }
\affil{Department of Astronomy; University of
Texas; Austin, TX 78712-1083
\\ dll@astro.as.utexas.edu}

\author{Guillermo Gonzalez}
\affil{Department of Astronomy;
University of Washington;
Seattle, WA 98195-1580\\gonzalez@astro.washington.edu}

\author{Gajendra Pandey   }
\affil{Department of Astronomy; University of
Texas; Austin, TX 78712-1083
\\ pandey@astro.as.utexas.edu}

\begin{abstract}

 We present  the chemical compositions of four K giants CS 22877-1,
 CS 22166-16, CS22169-35 and BS 16085 - 0050  that have [Fe/H] in
the range  $-$2.4 to $-$3.1. Metal-poor stars with [Fe/H] $<$ $-$2.5 are
 known to exhibit considerable star - to - star variations of many
 elements.
This quartet confirms this conclusion.
 CS 22877-1 and CS 22166-16 are carbon-rich.
 There is significant spread for [$\alpha$/Fe] within our sample where
[$\alpha$/Fe] is computed from the mean of the [Mg/Fe],  and [Ca/Fe]
ratios. 
 BS 16085 - 0050 is remarkably $\alpha$ enriched with a mean [$\alpha$/Fe]
 of $+$0.7 but CS 22169-35 is $\alpha$-poor.
 The aluminium abundance also shows
a   significant variation over the sample.
A parallel and unsuccessful
 search among high-velocity late-type stars for metal-poor
stars is described.

{\it Subject headings: stars:abundances -- stars:chemically peculiar     
stars: late-type}
\end{abstract}

\section{Introduction}

Chemical compositions of metal-poor stars have long been used
to probe the history of the early Galaxy. As the number of very metal-poor
stars having a well determined chemical composition has increased,
it has become apparent that the metallicity, usually represented
by the iron abundance [Fe/H], is not sufficient to predict the abundances
of other elements; a real star-to-star scatter in
abundance ratios [el/Fe] appears for many elements among stars
more metal-poor than [Fe/H] of about -2. For  stars in [Fe/H] range
$-$2 to $+$0.3,                                                            
the various [el/Fe] show almost no `cosmic' scatter. In  the light of the
cosmic scatter shown by very metal-poor stars, it is important to analyse
as large a sample of such stars as possible, in order to characterize
fully the extent of the scatter. This paper represents a modest contribution
to that end by presenting abundance analyses of four K giants with
[Fe/H] $\sim$ -2.4 to -3.1, and by describing an as yet unsuccessful
search for late-type metal-poor stars in a list of stars of high tangential
velocity.

\section{Observations}
    
 A program of medium resolution spectroscopy was undertaken
 to identify metal-poor  field stars  from 
 Lee's (1984) list
of stars with tangential velocities estimated to exceed 100 km s$^{-1}$.
Lee compiled his lists  from
Lowell proper motions (Giclas, Burnham \& Thomas 1971) and an
estimate of the parallax (trigonometric or spectroscopic). 
We elected to concentrate on stars of spectral type
K because major surveys of
metal-poor stars have largely been restricted to earlier spectral
types (cf. Carney et al. 1996).
 Stars of spectral type K  provide
opportunities to measure aspects of a star's chemical composition
not readily determinable from warmer stars. 
The stars  observed with medium resolution are 
presented in  Table 1. Estimates of
the B or V  magnitude, and spectral type are taken from SIMBAD. The
radial velocity is measured from our spectra (see below). 
In addition to stars from Lee's sample, we observed 3 stars from the HK survey
undertaken by Beers, Preston, \& Schectman (1985, 1992).

 The  Cass\'{e}grain spectrograph 
at  the 2.3m Vainu Bappu Telescope (VBT) at 
Kavalur (Prabhu, Anupama \& Surendiranath 1998) was used for the
spectroscopic survey.
 With the chosen grating of 1200 grooves mm$^{-1}$ and a Tektronix
1024 $\times$ 1024 24$\mu$m pixel CCD, the  
intermediate dispersion was  1.3\AA~ per pixel.
 The bright K subgiant (spectral type K1IV)
 star HR~4182 was  observed  as radial velocity standard and also as 
a  near-solar metallicity representative. 

These  spectra enabled us to measure radial velocities (Table 1)
 to an accuracy of
about $\pm$ 18 km s$^{-1}$ using cross-correlation task FXCORR contained
in IRAF software.
Our velocities are in fair  agreement with the
few previously published values located through SIMBAD (see footnotes to
Table 1). 
 From inspection of the spectra, it was obvious that five stars from Table 1
 are  extremely
metal-poor: CS 22166-16, 22169-35, 22877-1, 22877-11, 22877-51.
A metal deficiency for these stars was expected from the preliminary
estimates of metallicity assigned by Beers et al. (1992).
CS 22877-11 was  the subject of a
detailed abundance analysis reported by McWilliam et al. (1995) who
gave the iron deficiency as [Fe/H] = -2.9. Str\"{o}mgren
photometry (Schuster et al. 1996) showed CS 22877-51 to be
metal-poor ([Fe/H] = -2.45).   
Lee's star G 53-24 is given [Fe/H] = 0.18 by
Ryan \& Norris (1991).

The three  very metal-poor stars from the Kavalur survey
with  BS 16085-0050 from Anthony-Twarog et al. (2000)
 were selected for an abundance analysis. 
They were observed
 with the Apache Point Observatory's 3.5 meter telescope and the
vacuum-sealed echelle spectrograph.  It uses a 31.6 line/mm echelle 
grating with a prism cross-disperser.  The 2048x2048 SITe CCD has 
24 micron pixels, resulting in a 2 pixel resolving power near 38,000.
Due to the close spacing of the orders on the CCD, it is necessary to 
employ nonstandard reduction methods.  The most significant difference 
is the use of a hot star instead of lamp as a flat.
As a measure of the quality of the spectra, we note that the
S/N  ratio (per pixel) in the continuum near 6700\AA\ 
 was about 70 (CS 22166-16),  60  (CS 22169-35),
 75  (CS 22877-1), and  90  (BS 16085-0050).

\section{Abundance Analysis}

Inspection of the high-resolution spectra confirmed that the 
observed stars are very metal-poor.
 Hydrogen lines appeared normal with no indication
of emission. 
Our abundance analysis was an entirely standard procedure, as
described for example in our papers on the RV Tauri variables
(cf. Giridhar, Lambert \& Gonzalez 2000). A 1997 version of the spectrum
synthesis code MOOG (Sneden 1973) was used with model atmospheres
drawn from the grid computed by Kurucz (1993). Lines of Fe\,{\sc i} and
Fe\,{\sc ii} were used to derive the atmospheric parameters: effective
temperature T$_{\rm eff}$, surface gravity $g$, and the microturbulent
velocity $\xi_t$. The requirement  that the Fe abundance derived from 
Fe\,{\sc i} lines be independent of excitation potential and equivalent
width were used to derive T$_{\rm eff}$ and $\xi_t$, respectively. Then,
the requirement that Fe\,{\sc i} and Fe\,{\sc ii} lines return the
same abundance was used to derive $\log g$.  This use of ionization
equilibrium was verified using the Ti\,{\sc i} and Ti\,{\sc ii}
lines. Given the number of lines and the
accuracy of the equivalent widths (5-10\%), we consider the uncertainties
for these quite similar stars to be about $\pm 150$K in T$_{\rm eff}$,
$\pm$  0.25  in $\log g$, and $\pm 0.2$ km s$^{-1}$ in $\xi_t$. The
iron abundance is determined to about $\pm$ 0.2 dex. Final results
for the atmospheric parameters are given in Table 2. Inspection of
the abundances derived for the individual stars (Table 3) show that
ionization equilibrium of Ti and Fe is satisfied with very similar
parameters; the abundance differences from Ti\,{\sc i} and Ti\,{\sc ii}
lines are equal to those from the Fe\,{\sc i} and Fe\,{\sc ii} lines to
within 0.15 dex for all stars.   

Lines  employed in this analysis are
 presented in the Table 4.
 The $\log gf$-values are taken
from compilations provided by  Luck (private communication)
and McWilliam et al. (1995).
 We derived carbon abundance for CS 22877-1, CS 22166-16 and CS 22169-35
 using CH band at 4310 to 4330\AA~. No CH line could be detected for
 BS 16085-0050 inspite of its low temperature.
 We used HFS parameters given in McWilliam et al. for Mn I lines
 at 4030,4033 and 4034\AA~ to synthesize this region for BS 16085-0050 
 to derive Mn abundance.
 The same could not be done  for CS 22877-1, CS 22166-16 and CS 22169-35
 as Mn I lines were contaminated by CH lines. 
Elemental abundances are given in Table 3    
as [X/H] with the standard error  derived from the line-to
-line scatter, the number of adopted lines, and [X/Fe].  Uncertainties
arising from likely errors in the atmospheric parameters can be assessed
from Ryan, Norris \& Beers (1996, Table 3 and entries for HD~122563). In almost
all cases, our tabulated standard error is the dominant contributor to the
total error; the standard error of the mean is formally smaller than
that quoted by the square-root of the number of contributing lines. 

\section{Discussion}

In the light of published results  on the composition of extremely
metal-poor stars, we commence our discussion by comparing and
contrasting the  four stars with  previously
analysed stars. This will be done  using the ratio
[X/Fe]. Our primary reference works are the  surveys
by McWilliam et al. (1995) and Ryan et al. (1996), and reviews by
McWilliam (1997) and Norris (1999). Variation of [X/Fe] with [Fe/H] is well
determined from [Fe/H] $\sim 0$ to
[Fe/H] $\simeq$ -2 with remarkably little true (cosmic)
scatter as long as normal stars are considered (Lambert 1989;
Wheeler, Sneden \& Truran 1989). For [Fe/H] $\leq -2$,  many
relations change shape and slope and may develop a significant cosmic
scatter. Our stars will be judged as normal if they fall within the
range of [X/Fe] given in the reference works. It should be noted that
the analytical tools (models, lines, etc.)
used here are very similar to those employed
by McWilliam et al. and Ryan et al. and, both samples included not only
dwarfs but
giants  similar to our stars. Therefore,   systematic
errors affecting our [X/Fe] should be similar 
to those of the reference works. Since our principal goal is to
relate our stars to the  previously analysed stars, we do not here
concern ourselves with the systematic errors arising from defects in
the analytical tools, e.g., the use of local thermodynamic equilibrium (LTE)
when non-LTE effects may be  significant.

\subsection{CS 22877-1}

This C-rich star at [Fe/H] = -2.8 has [X/Fe] firmly within the expected range
except for  Na (possibly underabundant), Al,
 and  Ti (possibly slightly underabundant).
 The high C abundance is
evident from the great strength of the CH bands:
our estimate of [C/Fe] = +1.8 comes from a spectrum synthesis (Figure 1).
Although this star is C-rich relative to most extremely
metal-poor stars, its [C/Fe] is matched by other stars (Norris, Ryan \&
Beers 1997).
 The Na abundance, [Na/Fe] = -0.5, is a little outside the range
of -0.3 to +0.4 reported by McWilliam et al., also using the Na D lines.
Figure 2 shows the Al I line at 3961\AA~.
The Al abundance, [Al/Fe] = +0.2, is within the large range of
-1.0 to +0.5 found by McWilliam et al. from the 3961\AA\  Al\,{\sc i} resonance
line; the other resonance line at 3944\AA\ 
 was shown by Arpigny \& Magain (1983)
to be blended with several CH lines, even in stars where CH was not the
prominent feature of the spectrum that it is for CS 22877-1.
  Ryan et al.'s analyses of the resonance line, however, gave
a rather well defined
`plateau' at [Al/Fe] $\simeq$ -0.8 and relative to this value (and to
points in their [Al/Mg] vs  [Fe/H] plot),
CS 22877-1 is Al-enriched but analysis of the 3961\AA\ line is
sensitive to the adopted microturbulence. The origin of the difference between 
McWilliam et al.'s and Ryan et al.'s Al abundances 
is unclear, according to Ryan et  al. Clearly, there is a need to
probe this discrepancy more deeply because classification of
Al abundances for the four stars as `normal' or abnormal' depends
on whether the appropriate set of reference abundances is that
offered by McWilliam et al. or that by Ryan et al. We do not attempt to
resolve the discrepancy. 
 In Figure 3, we show the star's location
in [X/Fe] versus [Fe/H] plots for X = Al, $\alpha$,  where $\alpha$
denotes Mg, and Ca. Abundances from McWilliam et al. and Ryan et al.
are given as a reference.
 We chose not to include [Si/Fe] as not many good Si I lines were accessible.
 The abundances  of the three
measured elements from Sr to Ba  are each within the ranges defined by 
McWilliam et al. and Ryan et al. 

\subsection{CS 22166-16}

CS 22166-16 at [Fe/H] = -2.4 is enriched in carbon relative to the typical
value [C/Fe] $\simeq$ 0  but to a more moderate
and seemingly more common level than CS 22877-1. Other elements
are at or close to their expected levels. Mg and Ca abundances are 
well determined and provide [$\alpha$/Fe] = 0.6. The Si abundance
based on a single weak line is [Si/Fe] = 0.2.
Chromium is underabundant at [Cr/Fe] = -0.9; McWilliam et al. and Ryan et al.
find [Cr/Fe] $\simeq$ -0.2 for [Fe/H] = -2.4 with a few outliers at
lower [Cr/Fe].
Na has a normal abundance according to McWilliam et al.'s
results.
 The abundances of Sr and
Ba, the only measured heavy elements, fall within the cosmic scatter
previously reported with [Sr/Fe] at the lower boundary of previous
results.

\subsection{CS 22169-35}

Relative to the expected composition of a [Fe/H] = -2.9 star, a
striking aspect of CS 22169-35 is the low abundance of the
$\alpha$-elements Mg, Si, Ca, and Ti. Magnesium and calcium which are
well represented give [Mg/Fe] = 0.0 and [Ca/Fe] = 0.1. Titanium
from 9 Ti\,{\sc ii} lines gives [Ti/Fe] = 0.0 (Figure 3), a result
consistent with that from the 2 Ti\,{\sc i} lines.
The mean index [$\alpha$/Fe] from Mg and Ca (Figure 3) is smaller than
any index measured by McWilliam et al. and Ryan et al.  
 McWilliam et al. found
3 stars with a [Mg/Fe] well below the expected values, and one of these
stars was also low in Ca, the only low Ca star in their sample.
Carbon is slightly enriched: [C/Fe] $\simeq$ +0.4  where
[C/Fe] $\simeq$ 0.0 is accepted as normal.
Aluminium at [Al/Fe] = -0.7 is not exceptional and belongs to Ryan et al.'s
plateau. Sodium  from the Na D lines
appears distinctly underabundant relative to McWilliam et al.'s
sample at [Na/Fe] = -0.7.
 
 A metal-poor star with low $\alpha$-element abundances
while unusual is not unprecedented.
 Our star is reminiscent of  the pair
HD~134439/134440 (King 1997; James 2000) at [Fe/H] = -1.5. These common proper
motion  stars have a higher Ba abundance ([Ba/Fe] $\simeq$ -0.2 vs -1.1)
 but this may
reflect the metallicity difference  of 1.4 dex. At a metallicity closer to
that of CS 22169-35, BD~+$80^\circ$\,245 (Carney et al. 1997; James 2000)
with [Fe/H] = -2.0 has low abundances of the $\alpha$-elements (Mg, Ca, and
Ti at [$\alpha$/Fe] $\simeq$ -0.2) and  Ba  ([Ba/Fe] = -1.3). 
James also found low Mg, and Ca (and Ba) but not Ti abundances in
the star G 4-36 with
[Fe/H] = -2.0.
A group of  $\alpha$-poor stars with [Fe/H] = -1.2 to -0.6
 was uncovered by Nissen
\& Schuster (1997) who  delineated
several abundance trends. Their correlation between [Na/Mg] and [Mg/H] 
does not extend to lower [Fe/H] but our result, also the results by
James (2000), suggests an approximately constant [Na/Mg] at low
[Fe/H]. On the other hand, the [Ni/Fe] vs [Fe/H] trend suggested
by Nissen \& Schuster is not satisfied by $\alpha$-poor low [Fe/H] stars.

\subsection{BS 16085-0050}

The outstanding aspect of this star's composition,
  the most iron-deficient  of the quartet at [Fe/H] = -3.1,
is the high abundance of
the $\alpha$-elements. With [$\alpha$/Fe] = 0.9, 1.2, and 0.6 for Mg, Si,
and Ca, respectively, it is unmatched by any of Ryan et al.'s stars.
McWilliam et al. found one star with similarly high [$\alpha$/Fe]
values: CS  22949-037 with [$\alpha$/Fe]  = +1.2, 0.9, and 0.9 for
Mg, Si, and Ca, respectively, and [Fe/H] = -4.0. Titanium sometimes
grouped with the $\alpha$-elements
is a little below its expected abundance  at [Ti/Fe] = 0.2.
In CS 22949-037, Ti is only marginally above its expected value.

BS\,16085-0050's high abundance of $\alpha$-elements is a robust
result. 
A high [$\alpha$/Fe] is clearly indicated for Mg and Ca. Magnesium is
represented by 7 Mg\,{\sc i} lines including 2 strong lines and 5 weaker
lines. Seven medium-strong Ca\,{\sc i} lines give consistent results. If the
atmospheric parameters are varied by their estimated uncertainties [Mg/Fe] and
[Ca/Fe] do not vary by more than about $\pm$0.05. No plausible change can
reduce the best estimates [Mg/Fe] = 0.9 and [Ca/Fe] = 0.6 to the lower
more typical values. Silicon is represented by just 2 lines. Variation
of the atmospheric parameters has a larger effect on the [Si/Fe] ratio 
because in part the Si\,{\sc i} lines are sensitive to the adopted
microturbulence but the effect is much smaller than the difference
between the estimated [Si/Fe] = 1.2 and the lower value from the published
surveys. Titanium for which adequate samples of Ti\,{\sc i} and
Ti\,{\sc ii} lines provide consistent results clearly gives a lower ratio
of [Ti/Fe] = 0.2 with very little sensitivity to the atmospheric
parameters. In summary, Mg and Ca and most likely Si but not Ti
are unusually enriched in
this star. Magnesium and calcium are well represented in all stars and,
therefore, the star-to-star differences in [$\alpha$/Fe] are considered
real.

Aluminum
is approximately normal judged by McWilliam et al.'s range in [Al/Fe],
and by  Ryan et al.'s [Al/Mg] but not their [Al/Fe] range where BS 16085-0050
appears overabudant in Al. 
Heavy elements Sr, Y, and Ba fall just within the previously reported
spreads for stars with [Fe/H] $\sim$ -3.  Their [X/Fe] values are
at the lower boundaries of previous results.

\section{Concluding Remarks}

A somehwat surprising outcome of our initial survery of Lee's
high velocity stars is the absence of metal-poor stars. Most
stars appear to have an abundance near solar. Given that the selection
criterion was a tangential velocity in excess of 100 km s$^{-1}$, we
expected to find a rather metal-poor sample. This apparent puzzle
will be considered when our survey is more complete.

Our modest addition to the number of very metal-poor stars subjected to
an abundance analysis reveals several interesting facets about these stars. 
Two of the stars are evidently very rich in carbon.
This is not an uncommon feature of very metal-poor
stars (Rossi, Beers \& Sneden 1999).
 At solar metallicities, carbon-enrichment of a stellar
atmosphere is widely associated with enrichment of $s$-process
elements and
attributed to contamination of the star by mass-transfer from an AGB star.
This attribution is far less appropriate for very metal-poor stars.
Several C-enriched stars have been shown
not to be binaries (Norris et al. 1997) and in addition the C-enrichment is not
always coupled with $s$-process enrichment. In short, mass transfer
is a plausible explanation in only some cases and in others the C-enrichment
was likely present in the star's natal cloud. In the case of CS 22877-1 and
probably also CS 22877-16, the carbon enrichment is not related to the
$s$-process but heavy elements point to an $r$-process connection.
Whether these stars are binaries or not is presently unknown.

Very metal-poor stars are apparently
not a monolithic block with respect to the ratio of
$\alpha$-elements to iron. BS 16085-0050 is  $\alpha$-enriched 
at a level rarely seen previously but this enrichment is
not unprecedented:  CS 22949--37 is similar (McWilliam et al. 1995).
At the other extreme, CS 22169-35 joins a small group of $\alpha$-poor
very metal-poor stars. Star-to-star spread in elemental abundance ratios
with respect to iron are seen in some other elements, particularly the
heavy elements where a considerable dispersion has been noted
previously.

The large spread in abundances particularly for $\alpha$ elements 
 found for stars with [Fe/H] $<$ $-$2.5 indicate that before this 
 metallicity level was attained, the early Galaxy experienced very 
 unusual chemical enrichment. Refinement of supernovae models are
 perhaps required to explain uneven yield of $\alpha$ elements.
 Galactic chemical enrichment models including the effect of 
 incomplete mixing might help in explaining the observed abundance
 peculiarities.
 Perhaps, the principal lesson to be drawn from our
small sample is that not everything has yet been gleaned from
abundance analyses of very metal-poor stars. The mine is not yet
exhausted of valuable ores.

At the University of Texas, this research was supported in part by the
National Science Foundation (grant AST-9618414) and the Robert A. Welch
Foundation.

\section{references}

Anthony-Twarog, B.J., Sarajedini, A., Twarog, B.A., \& Beers, T.C.
   2000, AJ, 119, 2882

Arpigny, C., \& Magain, P. 1983, A\&A, 127, L7




Beers, T.C., Preston, G.W., \& Schectman, S.A. 1985, AJ, 90, 2089

Beers, T.C., Preston, G.W., \& Schectman, S.A. 1992, AJ, 103, 1987


Carney, B.W., Laird, J.B., Latham, D.W., \& Aquilar, L.A. 1996, AJ, 112, 668 

Carney, B.W., Wright, J.S., Sneden, C., Laird, J.B., Aguilar, L.A.,
  \& Latham, D.W. 1997, AJ,114, 363

Giclas, H.L., Burnham, R. Jr., \& Thomas, N.G., 1971, Lowell Proper
 Motion Survey - Northern Hemisphere, Lowell Observatory.

Giridhar, S., Lambert, D.L., \& Gonzalez, G. 2000, ApJ, 531, 521

James, C.R. 2000, PhD Thesis, Univ. of Texas at Austin

King, J.R. 1997, AJ, 113, 2302

Kurucz, R.L., 1993, ATLAS9 Stellar Atmosphere Programs and 2 km s$^{-1}$
  grid, CDROM 13, Smithsonian Astrophysical Observatory

Lambert, D.L. 1989, in {\it Cosmic Abundances of Matter}, ed. C.J. Waddington,
   AIP Conf. Proc. 183, 168

L\`{e}bre, A., de Laverny, P., de Medeiros, J.R., Charbonnel, C., da Silva, L.
 1999, A\&A, 345, 936   

Lee, S.-G. 1984, AJ, 89, 720.;

McWilliam, A. 1997, ARAA, 35, 503 

McWilliam, A., Preston, G.W., Sneden, C., \& Searle, L. 1995, AJ, 109, 2757

Nissen, P.E., \& Schuster, W.J. 1997, A\&A, 326, 751

Norris, J.E. 1999, in {\it The Third Stromlo Symposium: The Galactic Halo},
  ed. B.K. Gibson, T.S. Axelrod, \& M.E. Putman, ASP Conf. Ser., 165, 213

Norris, J.E., Ryan, S.G., \& Beers, T.C. 1997, ApJ, 488, 350 

Prabhu, T.P., Anupama, G.C., \& Surendiranath, R. 1998, BASI, 26, 383

Rossi, S., Beers, T.C., \& Sneden, C.
  1999, in {\it The Third Stromlo Symposium: The Galactic Halo},
  ed. B.K. Gibson, T.S. Axelrod, \& M.E. Putman, ASP Conf. Ser., 165, 264

Ryan, S.G., \& Norris, J.E. 1991, AJ, 101, 1835

Ryan, S.G., Norris, J.E., \& Beers, T.C. 1996, ApJ, 471, 254

Schuster, W.J., Nissen, P.E., Parrao, L., Beers, T.C., \& Overgaard, L.P.
   1996, A\&AS, 117, 317

Sneden C. 1973, PhD Thesis, Univ. of Texas


Wheeler, J.C., Sneden, C., \& Truran, J.W. 1989, ARAA, 27, 279

\section{Figure Captions}

\figcaption{Observed and synthetic spectra of CS 22877-1 at high resolution
showing CH lines. The synthetic spectrum is computed for [C/Fe] = +1.8.}

\figcaption{Observed spectra of CS 22877-1  and BS 16085 - 0050    
 showing Al I line at 3961\AA.  }

\figcaption{[X/Fe] vs [Fe/H] for  X= Al, $\alpha$ and Mn where [$\alpha$/Fe]
is the mean of [Mg/Fe] and [Ca/Fe].
Abundances from McWilliam et al. and Ryan et al. are
 given as reference.}
 
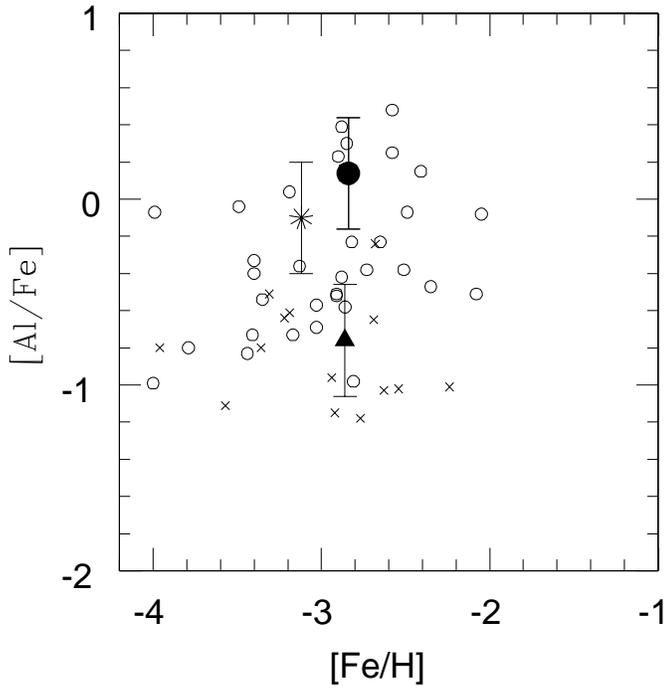
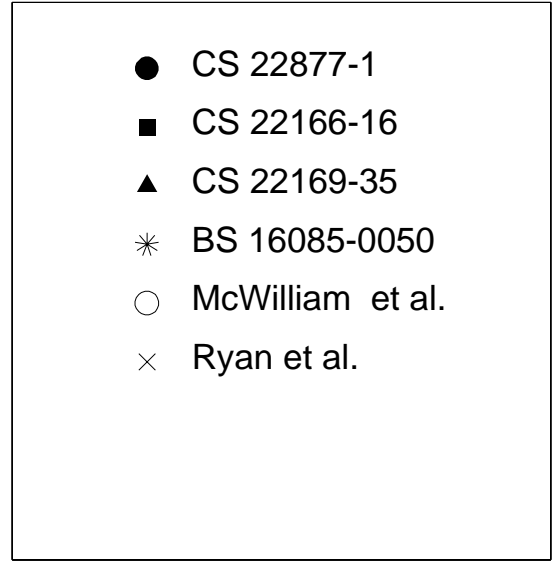
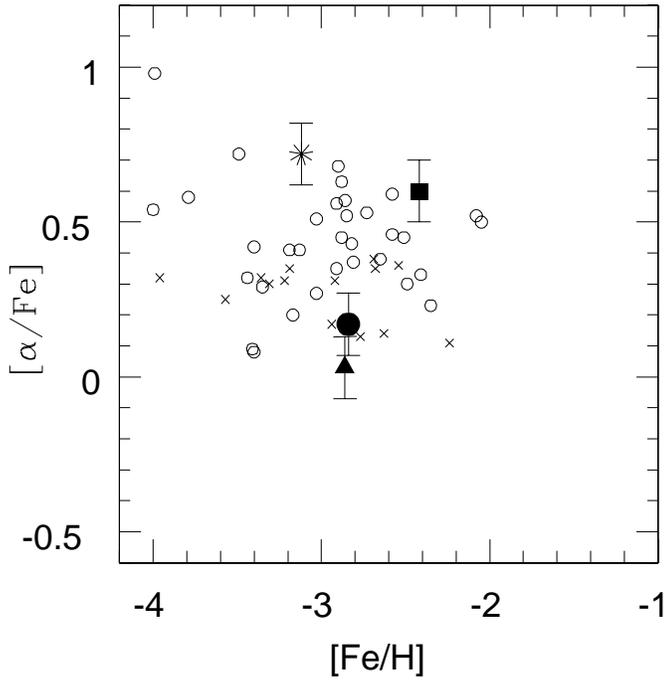
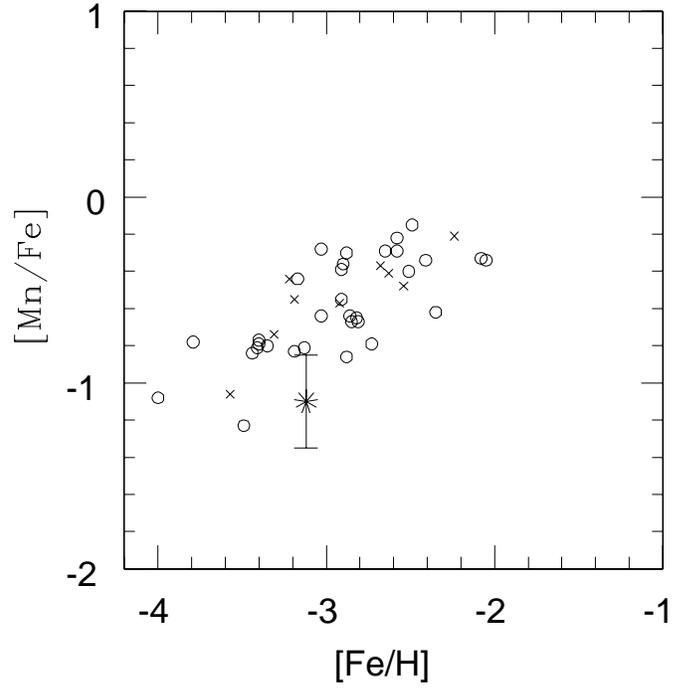
\end{document}